\documentclass[]{aa}\usepackage{graphics,epsf}
\newcommand{\lta}{\;
  \raise0.3ex\hbox{$<$\kern-0.75em\raise-1.1ex\hbox{$\sim$
  }}\;\hskip-2pt }
\newcommand{\gta}{\;
  \raise0.3ex\hbox{$>$\kern-0.75em\raise-1.1ex\hbox{$\sim$
  }}\;\hskip-2pt }
\begin{document}
\title{Effects of boundary conditions on the dynamics of  the solar convection zone}
\author{Reza Tavakol\thanks{e-mail: r.tavakol@qmul.ac.uk}\inst{1}
\and Eurico Covas\thanks{e-mail: e.o.covas@qmul.ac.uk}\inst{1}
\and David Moss\thanks{e-mail: moss@ma.man.ac.uk}\inst{2}
\and Andrew Tworkowski\thanks{e-mail: a.s.tworkowski@qmul.ac.uk}\inst{3}}
\institute{Astronomy Unit, School of Mathematical Sciences,
Queen Mary, University of London, Mile End Road, London E1 4NS, UK
\and Department of Mathematics, The University, Manchester M13 9PL, UK
\and Mathematics Research Centre, School of Mathematical Sciences,
Queen Mary, University of London, Mile End Road, London E1 4NS, UK
}
\date{Received ~~ ; accepted ~~ }
\offprints{\em R.\ Tavakol}
\titlerunning{Tavakol et al.: Effects of boundary conditions on the dynamics of solar convection zone}
\markboth{Tavakol et al.: Effects of boundary conditions on the dynamics of solar convection zone}
{Tavakol et al.: Effects of boundary conditions on the dynamics of solar convection zone}
\abstract{
 Recent analyses of the helioseismic
data have produced evidence for a 
variety of interesting dynamical 
behaviour associated
with torsional oscillations.
What is not so far clear is whether these
oscillations extend all the way to the bottom of
the convection zone and, if so, whether 
the oscillatory behaviour 
at the top and the bottom of the
convection zone is different.
Attempts have been made to understand 
such modes of behaviour
within the framework of nonlinear dynamo models 
which include the nonlinear action
of the Lorentz force of the
dynamo generated magnetic field on the solar
angular velocity.
One aspect of these models that remains uncertain
is the nature of the boundary conditions on the magnetic field.
Here by employing a range of physically plausible 
boundary conditions, we show that
for near-critical and moderately supercritical 
dynamo regimes, the oscillations extend all 
the way down to the bottom of the
convection zone. Thus, such penetration is an extremely robust feature
of the models considered.
We also find parameter ranges for which the supercritical
models show spatiotemporal fragmentation
for a range of choices of boundary conditions.
Given their observational importance, 
we also make a comparative study of the amplitude
of torsional oscillations as a function of the boundary conditions.

\keywords{Sun: magnetic fields -- torsional oscillations --  activity -- spatiotemporal fragmentation}
}
\maketitle
\section{Introduction}
Recent analyses of the helioseismic data, both from the
Michelson
Doppler Imager (MDI) instrument on board the SOHO
spacecraft (Howe et al.\ 2000a) and the Global 
Oscillation Network Group (GONG)
project (Antia \& Basu 2000),
have provided strong evidence that
the previously observed
solar torsional oscillations (e.g.\ Howard \& LaBonte 1980;
Snodgrass, Howard \& Webster 1985;
Kosovichev \& Schou 1997; Schou et al.\ 1998), with
periods of about 11 years,
penetrate into the convection zone (CZ)
to depths of at least 10 percent in radius.

These studies have also produced rather conflicting
results concerning the dynamical behaviour 
near the bottom of the convection zone.
Thus Howe et al.\ (2000b)
find evidence for the presence of
torsional oscillations near the tachocline situated close to
the bottom of the convection
zone, but with a markedly shorter period of about $1.3$ years, whereas
Antia \& Basu (2000) do not find such oscillations.
Given the uncertainties in the helioseismic data,
what is not certain so far is (i)
whether torsional oscillations do extend all the way
to the bottom of the CZ
and (ii) whether there are
different oscillatory modes of behaviour 
at the top and the bottom of the
CZ. 

Work is in progress by a number of groups to
repeat these analyses in order 
to answer these observational questions.
In parallel, attempts have been made to approach these
questions theoretically by modelling
variations in the CZ
within the framework of nonlinear dynamo models
which include a nonlinear action
of the azimuthal component of the Lorentz force of the
dynamo generated magnetic field on the solar
angular velocity (Covas et al.\ 2000a,b; Covas et al.\ 2001a,b,
see also erratum in Covas et al. 2002).
According to these results, for most ranges of 
dynamo parameters, such as the dynamo and Prantdl numbers,
the torsional oscillations 
extend all the way down to the bottom of the convection 
zone. 
In addition,
{\it spatiotemporal fragmentation/bifurcation} (STF)
has been
proposed as a dynamical mechanism to
account for the possible existence of multi-mode
behaviour in different parts of the
solar CZ (Covas et al.\ 2000b, 
2001a,b, 2002).
In all these studies the underlying zero order angular velocity was
chosen to be consistent with
the recent helioseismic data.

As in much
astrophysical modelling,
an important source of uncertainty in these models is
the nature of their boundary conditions.
Given this uncertainty, and the fact that 
boundary conditions can alter qualitatively the behaviour of
dynamical systems,
it is important  to
see whether employing different
boundary conditions can significantly change
the dynamics in the CZ, and in particular
whether the two dynamical modes of behaviour
mentioned above 
are robust with respect to plausible changes in the boundary conditions.
This is important for two reasons. Firstly, in the absence of
precise knowledge about such boundary conditions,
it is important that the dynamical phenomena of interest predicted
by such models can survive
reasonable changes in ill-known boundary conditions.
Secondly, it may in principle be possible for qualitative changes 
found as the boundary conditions are altered 
to be used
as a diagnostic tool to determine the
range of physically reasonable boundary conditions in the
solar context.

Here, by considering a number of families of 
boundary conditions, we show that the penetration of
torsional oscillations to the bottom
of the CZ is indeed robust with respect
to a number of plausible changes to the boundary conditions.
We also find spatiotemporal fragmentation in these models
with a variety of, but not all, choices of boundary conditions.
Given the observational importance of the
amplitudes of the torsional oscillations, 
we also make a comparative study of their magnitudes
as a function of the boundary conditions.

\section{The model}
We shall assume that the gross features of the
large scale solar magnetic field
can be described by a mean field dynamo
model, with the standard equation
\begin{equation}
\frac{\partial{\vec B}}{\partial t}=\nabla\times({\vec u}\times {\vec B}+\alpha{\vec
B}-\eta\nabla\times{\vec B}).
\label{mfe}
\end{equation}
Here ${\vec u}=v\mathbf{\hat\phi}-\frac{1}{2}\nabla\eta$,
the term proportional to
$\nabla\eta$ represents the effects of turbulent diamagnetism,
and the velocity field is taken to be of the form $
v=v_0+v'$,
where $v_0=\Omega_0 r \sin\theta$, $\Omega_0$ is a prescribed
underlying rotation law and the component $v'$ satisfies
\begin{equation}
\frac{\partial v'}{\partial t}=\frac{(\nabla\times{\vec B})\times{\vec B}}{\mu_0\rho
r \sin\theta} . \mathbf{\hat {\vec \phi}}  + \nu D^2 v',
\label{NS}
\end{equation}
where $D^2$ is the operator
$\frac{\partial^2}{\partial r^2}+\frac{2}{r}\frac{\partial}{\partial r}+\frac{1}
{r^2\sin\theta}(\frac{\partial}{\partial\theta}(\sin\theta\frac{\partial}{\partial
\theta})-\frac{1}{\sin\theta})$  and $\mu_0$ is the induction constant.
The assumption of axisymmetry allows the field ${\vec B}$ to be split simply
into toroidal and poloidal parts,
${\vec B}={\vec B}_T+{\vec B}_P = B\hat\phi +\nabla\times A\hat\phi$,
and Eq. (\ref{mfe}) then yields two scalar equations for $A$ and $B$.
Nondimensionalizing in terms of the solar radius $R$ and time $R^2/\eta_0$,
where $\eta_0$ is the maximum value of $\eta$, and
putting $\Omega=\Omega^*\tilde\Omega$, $\alpha=\alpha_0\tilde\alpha$,
$\eta=\eta_0\tilde\eta$, ${\vec B}=B_0\tilde{\vec B}$ and $v'= \Omega^* R\tilde v'$,
results in a system of equations for $A,B$ and $v'$. The
dynamo parameters are $R_\alpha=\alpha_0R/\eta_0$, $R_\omega=\Omega^*R^2/\eta_0$,
$P_r=\nu_0/\eta_0$, and $\tilde\eta=\eta/\eta_0$, where $\Omega^*$
is the solar surface equatorial angular velocity.
Here
$\nu_0$ and $\eta_0$ are the turbulent magnetic
diffusivity and viscosity respectively
and $P_r$ is the turbulent Prandtl number.
Our computational procedure is to adjust $R_\omega$ so as to make the cycle
period be near the solar cycle period of about 22 years for the marginal
dynamo number, and then to allow $R_\alpha$ and, to some extent, $P_r$ to vary.
The density $\rho$ is assumed to be uniform.

Eqs. (\ref{mfe}) and (\ref{NS}) were solved using the code
described in Moss \& Brooke (2000) (see also
Covas et al. 2000b) together with the boundary conditions given below,
over the range $r_0\leq r\leq1$, $0\leq\theta\leq \pi$.
We set  $r_0=0.64$, and with
the solar CZ proper being thought to occupy the region $r \gta 0.7$,
the region $r_0 \leq r \lta 0.7$ can be thought of as an overshoot
region/tachocline.
In the following simulations we used a mesh resolution of $61 \times 101$
points, uniformly distributed in radius and latitude respectively.

In this investigation, we took $\Omega_0$ in
$0.64\leq r \leq 1$ to be given by an interpolation on the MDI data
obtained from 1996 to 1999 (Howe et al. 2000a).
We set  $\alpha=\alpha_r(r)f(\theta)$,
where $f(\theta)$ was chosen to be $\sin^2\theta\cos\theta$ or $\sin^4\theta\cos\theta$.
The angular structure of $\alpha$ is quite uncertain, and
both these forms have been used in the literature (see e.g. R\"udiger \& Brandenburg 1995)
and their choice here is simply 
to make the butterfly diagrams more realistic.
We took $\alpha_r=1$ in 
all or part
of the CZ (see below for details),
with cubic interpolation to zero at $r=r_0$ and $r=1$ in the 
cases where $\alpha_r\neq 1$ everywhere.
Throughout we take $\alpha_r\ge 0$
and $R_\alpha < 0$. Also, in
order to take some  account of the
likely decrease in the turbulent diffusion coefficient $\eta$
in the overshoot region, we allowed a simple
linear decrease from $\tilde\eta=1$ at $r=0.8$
to $\tilde\eta=0.5$ in $r<0.7$.


\section{The choice of boundary conditions}
Boundary conditions on magnetic fields are often rather ill--determined
when modelling astrophysical systems.
This is certainly true in the case
of the Sun and solar-type stars.
Given this uncertainty, we shall consider a
number of physically motivated families of boundary conditions and 
investigate the
consequences of each on the dynamics of the
CZ. In particular we shall study
whether they
allow penetration of torsional oscillations all the way 
to the bottom of the CZ as well as 
supporting spatiotemporal fragmentation.
We note that at 
$\theta=0$ and $\pi$
symmetry conditions imply $A=B=0$.
In this article we shall concentrate on the changes to the outer
boundary conditions only.

\subsection{Boundary conditions at $r=r_0$}
The detailed physics is uncertain near the base
of the computational region ($r=r_0$).
Given that the angular momentum flux out of  a
region with boundary ${\vec{S}}$ from the magnetic stresses is
$\int_{\vec{S}} (\vec{B} B r \sin \theta)d{\vec{S}}$,
we set 
$B=0$
on $r=r_0$ in order to ensure zero
angular momentum flux across the boundary and, correspondingly,
stress-free conditions were used for $v'$.
The condition $\partial A/\partial r=A/\delta$
crudely models $A$ falling to
zero at distance $\delta$ below $r=r_0$
(cf. Moss, Mestel \& Tayler 1990; Tworkowski et al.\ 1998). We
chose $\delta=0.03$, but the general nature of
the results is insensitive to this choice. Taking
$\delta>0$
is computationally helpful as it reduces somewhat
the field gradients near $r=r_0$, although it is
not essential.  

\subsection{The outer boundary conditions}
At the outer boundary $r=R$,
we shall, in view of the uncertainties regarding the
outer boundary conditions, consider a number of
different but physically reasonable
choices.

One of the common choices for the
outer boundary conditions adopted in literature
is the `vacuum' boundary condition, 
in which the poloidal field within $r=R$ is
smoothly joined, by a matrix multiplication, to an external vacuum solution;
the azimuthal field $B=0$.

Given the dynamic nature of the solar surface,
the usual vacuum conditions can, to some extent at least, be regarded
as a mathematically convenient idealization.
Some aspects of this issue have recently been 
discussed at length by Kitchatinov, Mazur \& Jardine
(2000), who derive `non-vacuum' boundary conditions on both
$B$ and ${\vec B}_P$.
We also consider families of boundary conditions
which deviate from the vacuum conditions and refer to these as `open'.
As a convenient and flexible general form for the boundary conditions 
at the surface, we write
\begin{equation}
\label{general}
r \frac{dA}{dr} + n_1 A =0,~~~~ 
r \frac{dB}{dr} + n_2 B =0
\end{equation}
where $n_1$ and $n_2$ are constants 
that parameterize the
boundary conditions and, to some extent, their degree of openness.
With $n_1 = 1$, $ n_2 =0$
the two conditions for $A$ and $B$ 
reduce to $\frac{d(rA)}{dr} = B_\theta = 0$ and
$\frac{dB}{dr} =0$ respectively.     
The condition $B_\theta=0$ has been adopted previously
by some investigators. The limit $n_2 \rightarrow \infty$ gives
the often used $B=0$.
As $n_1$ increases, the penetration
of the poloidal field through the surface decreases, and in the limit 
$n_1\rightarrow \infty$ the boundary condition is $A=0$, and all the poloidal
field lines then close beneath the surface $r=R$, which is thus the limiting field line.
Using the vacuum boundary condition for ${\vec B}_P$ gives poloidal field lines 
that mostly make a modest angle with the radial direction, and so
we can anticipate that, by taking small values of $n_1$ and large
values of $n_2$, we will obtain solutions that resemble in some ways those found by using
the vacuum boundary conditions mentioned above.
(But note that Eq.~(\ref{general}) gives strictly {\it local} conditions
on the field components, whereas the vacuum condition on the poloidal field is essentially
{\it nonlocal}.)

We note one further technical point. The angular momentum flux through $r=R$
is non-zero if both ${\vec B}_P$ and $B$ are non-zero there. Whilst the
Sun certainly is losing angular momentum, we are not trying to model this
process here, and so will only consider models in which the angular momentum
`drift' of the dynamo region is small enough that we can consider it to be 
a unchanging background for the dynamo calculations.

Now in order to find the range of values of  
$n_1$ and $n_2$ such that the resulting
`partially open' boundary conditions 
are physically plausible, we need to ensure that 
the chosen boundary conditions result in appreciable poloidal flux
penetrating the surface. 
Thus we calculated
the average over the dynamo cycle of the
ratio of the flux of the poloidal field at the surface to
the corresponding value within
the CZ, given by

\begin{equation}
\label{ratioA}
F_s  =\frac{[R \sin(\theta) A(R,\theta)]_{\rm Surface}}{[r \sin(\theta) A(r,\theta)]_{\rm Inner}},
\end{equation}
as a function of $n_1$ say.
Here $R$ is the model radius,
the numerator is evaluated at the surface $r=R$,
and the denominator is evaluated inside the
dynamo region (`CZ') $ r_0 \le r \le R $. As 
$n_1$ increases
we would expect this ratio to decrease. 
We consider a boundary condition to be, in principle, viable 
if the ratio $F_s$
is not too small compared with the corresponding value in the `standard'
vacuum case.

\section{Results}
Using the above model, we studied the dynamics in the convection zone subject to 
three sets of boundary conditions; namely, 
the vacuum boundary condition and two families of 
open boundary conditions which we shall refer
to as boundary conditions (1) and 
(2). Also in order to demonstrate that penetration of the 
torsional oscillations,
as well as spatial fragmentation,
can occur with various changes in other ingredients
of the model, we have chosen examples with different forms
for these ingredients.
\subsection {Vacuum boundary conditions} 

\begin{figure}
\centerline{\def\epsfsize#1#2{0.43#1}\epsffile{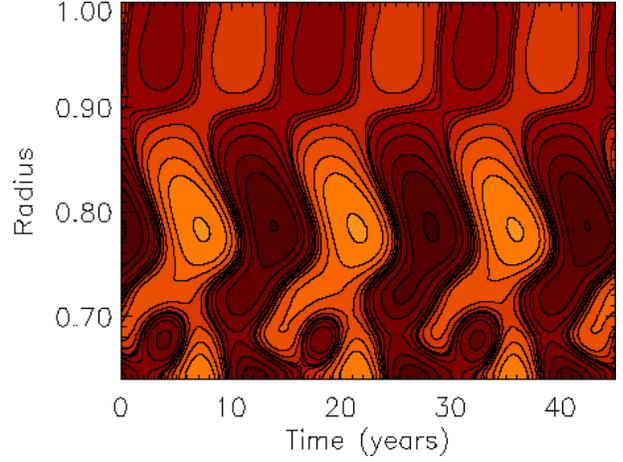}}
\caption[]{\label{ivac=1.velocity_radial_latitude=25.eps}
The radial (r--t) contours of the angular velocity 
residuals $\delta \Omega$ as a
function of time for a cut at $25$ degrees latitude, with
vacuum boundary conditions and
$R_\alpha = - 5.5, P_r= 0.6, R_\omega =60000$.
Note the fragmentation at the bottom of the
convection zone and the resulting
difference in periods of oscillations at the top and at the
bottom.
Darker and lighter regions represent positive and negative
deviations from the time averaged background rotation rate.
}
\end{figure}

\begin{figure}
\centerline{\def\epsfsize#1#2{0.44#1}\epsffile{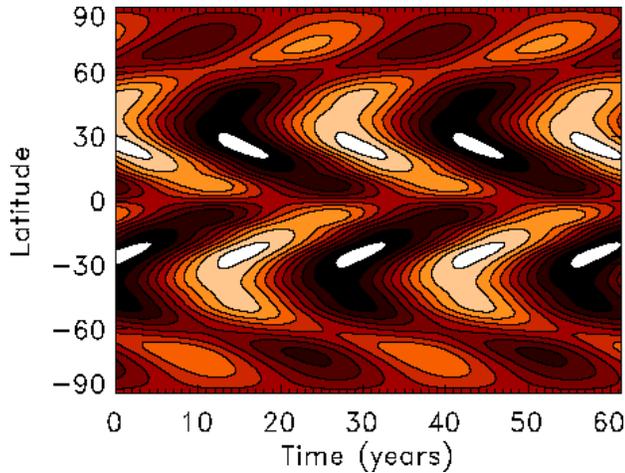}}
\caption[]{\label{ivac=1.butterfly_bp.R=0.90.eps}
Butterfly diagram of the toroidal component of the
magnetic field $\vec{B}$ at 
$r_0=0.90R$, with
the vacuum boundary conditions and the
parameter values given by
$R_\alpha = - 5.5, P_r= 0.6, R_\omega =60000$.
Dark and light shades correspond to positive and negative values
respectively.
}
\end{figure}

With this choice of boundary conditions, 
we found that for critical and moderately supercritical
regimes, the torsional oscillations extend all
the way down to the bottom of the
CZ (see also Covas et al. 2000a, where the
 critical value
of the dynamo number for the onset of dynamo was found to be
$R_\alpha \approx -3.16$).
In addition we found ranges of dynamo parameters
for which supercritical models
showed spatiotemporal fragmentation.
As an example of such fragmentation,
we show in
Fig.\ \ref{ivac=1.velocity_radial_latitude=25.eps} the radial 
contours of the angular velocity residuals $\delta \Omega$ as a
function of time for a cut at $25$ degrees latitude.
In this case we took $f(\theta)= \sin^4 \theta\cos\theta$,
with $\alpha_r=1$ throughout the computational region.
The parameter values used were
$R_\alpha = -5.5, P_r = 0.6$
and $R_\omega =60000$.
We also show in Fig. \ref{ivac=1.butterfly_bp.R=0.90.eps}
the magnetic butterfly diagram.

We note that this is the
first time STF has been obtained with vacuum
boundary conditions. This is despite the fact
that the range of parameter values
for which STF is present for this model is rather wide, 
as is the case with the  boundary conditions considered below.
What seems to occur here
is that the onset of
spatiotemporal fragmentation 
is close to a
bifurcation point, which
disrupts the butterfly diagram and rather confuses the diagnosis of the situation.
\subsection {Open boundary conditions (1)}
We now take boundary conditions at $r=R$ to be given by
Eq.~(\ref{general}), where
$n_2 =0$ and
$n_1$ is varied.

\begin{figure}
\centerline{\def\epsfsize#1#2{0.43#1}\epsffile{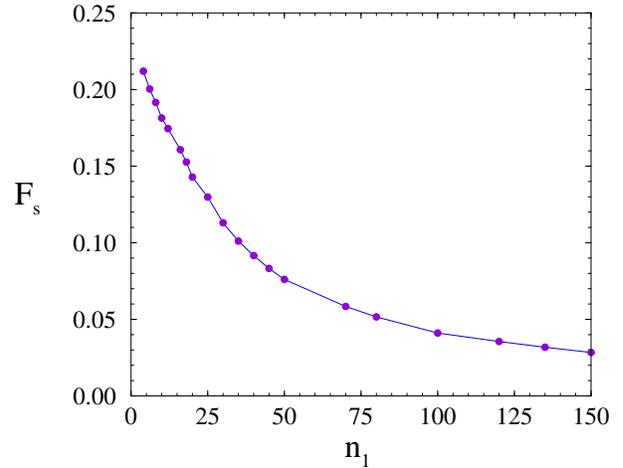}}
\caption[]{\label{amplitudes.flux.nfac.MDI.eps}
$F_s$ as a function of $n_1$ for the open boundary conditions (1)
given by equation (\ref{general}), with $n_2 =0$.
}
\end{figure}

In order to find the range of values of
$n_1$ which can be considered as physically plausible,
we calculated the ratio of the poloidal fluxes 
$F_s$ given by (\ref{ratioA}) as a function of
$n_1$. The results are given in Fig.\ \ref{amplitudes.flux.nfac.MDI.eps}.
To give an idea of how
this ratio compares to that of models with
vacuum boundary conditions, we also
calculated $F_s$ with vacuum boundary conditions with the  same
parameter values as for        
Fig.\ \ref{ivac=1.velocity_radial_latitude=25.eps}, and
found that $F_s \sim 0.25$. As can be seen from
Fig.\ \ref{amplitudes.flux.nfac.MDI.eps} this
is comparable in magnitude to the values obtained with the 
above open boundary
conditions for a wide range of values
of $n_1$ given by $n_1\lta 40$.

\begin{figure}
\centerline{\def\epsfsize#1#2{0.43#1}\epsffile{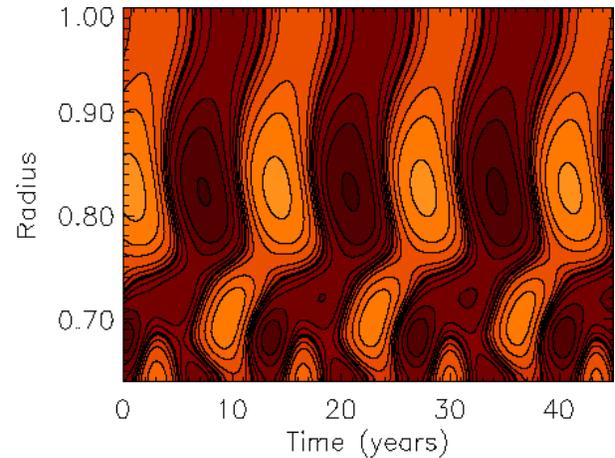}}
\caption[]{\label{ivac=7.velocity_radial_latitude=25.eps}
The radial (r--t) contours of the angular velocity 
residuals $\delta \Omega$ as a
function of time for a cut at $25$ degrees latitude, with
the boundary conditions given by (\ref{general}) with $n_1=25$,
$R_\alpha = -5.0, P_r = 0.7, R_\omega =50000$.
Note the fragmentation at the bottom of the
convection zone and the resulting
difference in periods of oscillations at the top and at the
bottom.
Darker and lighter regions represent positive and negative
deviations from the time averaged background rotation rate.
}
\end{figure}

\begin{figure}
\centerline{\def\epsfsize#1#2{0.44#1}\epsffile{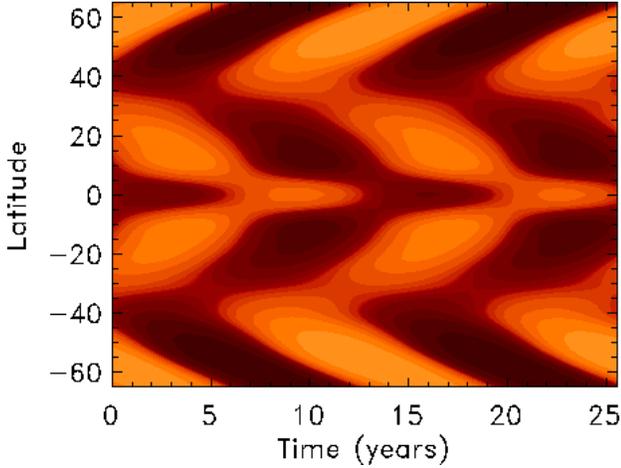}}
\caption[]{\label{ivac=7.velocity.R=0.92.eps}
Angular velocity residuals at $R=0.92$ with latitude and time,
with boundary conditions given by Eq.~(\ref{general}), $n_1=25$, 
$R_\alpha = -5.0, P_r = 0.7, R_\omega =50000$.
A temporal average has been subtracted to reveal the migrating banded zonal
flows. Darker and lighter regions represent positive and negative
deviations from the time averaged background rotation rate.
}
\end{figure}

\begin{figure}
\centerline{\def\epsfsize#1#2{0.38#1}\epsffile{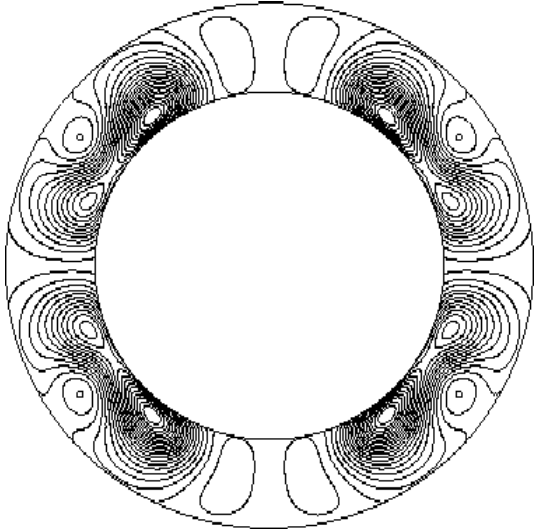}}
\caption[]{\label{ivac=7.a.r.sin.theta_snapshot_lines.eps}
The poloidal field lines;
same parameters as in Fig.\ \ref{ivac=7.velocity.R=0.92.eps}.
}
\end{figure}


\begin{figure}
\centerline{\def\epsfsize#1#2{0.38#1}\epsffile{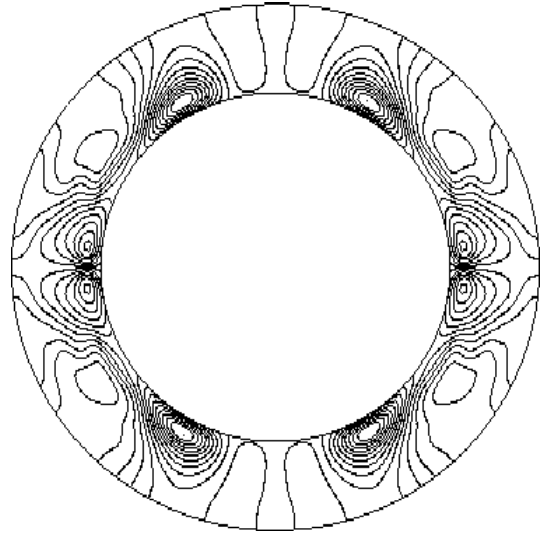}}
\caption[]{\label{ivac=7.bp_snapshot_lines.eps}
The toroidal field contours $\vec{B}_P$;
same parameters as for Fig.\ \ref{ivac=7.velocity.R=0.92.eps}.
}
\end{figure}

With these open boundary conditions,
we again found that
for slightly and moderately
supercritical
dynamo regimes, the torsional oscillations extend all
the way down to the bottom of the
CZ. There are also ranges of dynamo parameters
for which 
spatiotemporal fragmentation is found.
When $n_1$ is close to 1 there is
appreciable angular momentum drift, 
and so we do not consider these solutions further here.
As an example of a case with STF
(and with  negligible angular momentum
drift over the integration interval),
we show in Fig.\ \ref{ivac=7.velocity_radial_latitude=25.eps}
the radial
contours of the angular velocity residuals $\delta \Omega$ as a
function of time for a cut at $25$ degrees latitude.
In this case $f(\theta)= \sin^4 \theta\cos\theta$
and $\alpha_r$ is given by
$\alpha_r=1$ for $0.7 \leq r \leq 0.8$
with cubic interpolation to zero at $r=r_0$ and $r=1$.
The parameter values used are
$R_\alpha = -5.0, P_r = 0.7, R_\omega =50000$ and
$n_1=25$.

Fig.\ \ref{ivac=7.velocity.R=0.92.eps} shows the
angular velocity residuals at $R=0.92$
(i.e. the near-surface torsional oscillations),
with latitude and time,
for the same parameter values as in 
Fig.\ \ref{ivac=7.velocity_radial_latitude=25.eps}. 
The poloidal field lines and toroidal field contours for
this case are presented in Figs.\ \ref{ivac=7.a.r.sin.theta_snapshot_lines.eps} and 
\ref{ivac=7.bp_snapshot_lines.eps}.
In Fig.~\ref{ivac=7.bp_snapshot_lines.eps}, the effect of the open boundary condition
on $B$ is seen to be relatively minor -- most of the toroidal field is
concentrated near the tachocline, and far from the surface. The
poloidal field (Fig.~\ref{ivac=7.a.r.sin.theta_snapshot_lines.eps}) is
more uniformly distributed.
\subsection {Open boundary conditions (2)}
We now consider boundary conditions given by
Eq.~(\ref{general}), with 
$n_1$ of order one and
$1 \le n_2 \le 500$ 
(so $B\approx 0$ for large $n_2$).
We found that for $n_1$ values close to one
our model again has significant angular momentum drift,
which increases as
$n_1$ 
decreases to
$1$. For larger values of 
$n_1$ the drift is negligible, and the ratio $F_s$
is again `reasonable', being comparable with the
vacuum case.
We found that putting $n_1=2$ gave satisfactory behaviour, and this
is the case that we discuss in detail below.

\begin{figure}
\centerline{\def\epsfsize#1#2{0.43#1}\epsffile{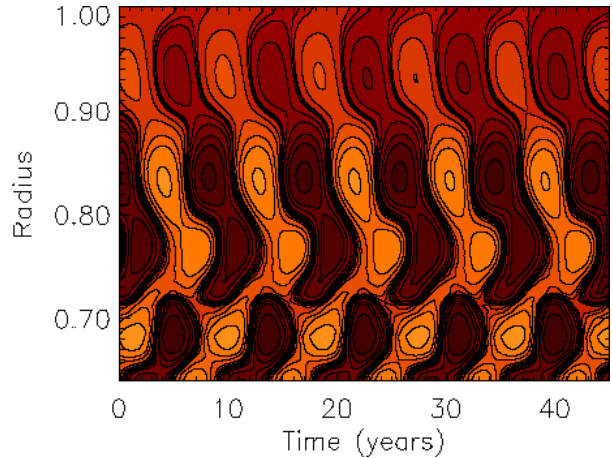}}
\caption[]{\label{ivac=9.velocity_radial_latitude=20.eps}
The radial (r--t) contours of the angular velocity 
residuals $\delta \Omega$ as a
function of time for a cut at $20$ degrees latitude, with
the open boundary conditions (2) and
$R_\alpha = -16, P_r= 0.22, R_\omega =48000$,
$n_1 =2$ and $n_2=400$.
Note the fragmentation at the bottom of the
convection zone and the resulting
difference in periods of oscillations at the top and a 
region near the bottom of the CZ around $r_0=0.7$.
Darker and lighter regions represent positive and negative
deviations from the time averaged background rotation rate.
}
\end{figure}

For all such cases we again found
that for slightly and moderately supercritical
dynamo numbers, the oscillations extend all
the way down to the bottom of the
CZ. In addition we found ranges of dynamo parameters
for which the supercritical model show spatiotemporal fragmentation.
An example, which has negligible angular momentum
drift over the integration interval,
is shown in Fig.\ \ref{ivac=9.velocity_radial_latitude=20.eps}.
Here $\alpha_r=1$ for $0.7 \leq r \leq 0.8$
with cubic interpolation to zero at $r=r_0$ and $r=1$ and 
$f(\theta)= \sin^2 \theta\cos\theta$.
The parameter values used were 
$R_\alpha = -16.0, P_r = 0.22$ and $R_\omega =48000$,
and boundary conditions were given by (\ref{general}) with $n_1 =2$ and $n_2=400$.

\subsection{Amplitudes of oscillations as a function of boundary conditions} 
Another important issue from 
an observational point of view is the
way the amplitudes of the 
torsional oscillations vary as a function of
model ingredients and parameters, as well as with depth
in the CZ.

To begin with, we verified that for 
a given boundary condition
the amplitudes of the oscillations  
increase as the dynamo number $R_\alpha$ and 
the Prantdl number $P_r$ are increased
(see also Covas et al 2001a). For orientation,
we recall that the observed surface amplitudes in the case
of the Sun are latitude dependent 
and of order of one nHz (see e.g. Howe et al.\ (2000b)).

We also made a comparative study of
the amplitudes as a function of changes
in the boundary conditions.
Briefly we found that typically the solutions with vacuum boundary conditions
have rather smaller amplitudes of oscillations, especially
near the surface.
For example, for the model of Fig.\ 1 (which has
a high $R_\alpha$, spatiotemporal fragmentation and
thus would be expected to have higher amplitudes),
we found the mean averaged amplitudes to be 
$0.72, 0.19, 0.09$ nHz
at the depths $r_0= 0.70, 0.88$ and $0.95$ respectively.

\begin{figure}
\centerline{\def\epsfsize#1#2{0.44#1}\epsffile{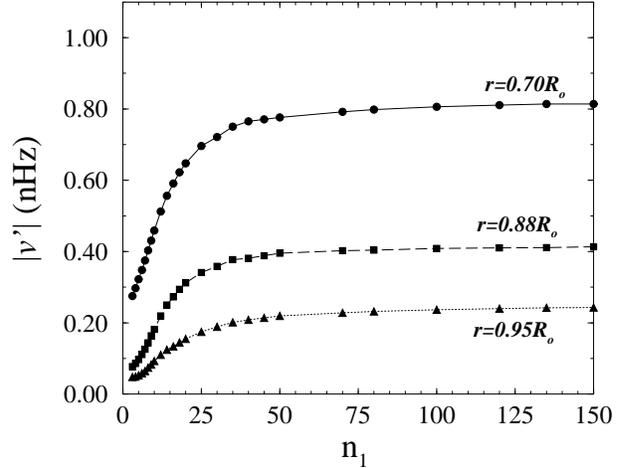}}
\caption[]{\label{amplitudes.nfacb.MDI.sin4.ivac=9.clean.eps}
The mean amplitudes of torsional oscillations,
at radii $r_0= 0.70, 0.88$ and $0.95$. in 
models with open boundary conditions given by
Eq.~(\ref{general}) and $n_2=0$, with $n_1$ ranging
from 3 to 150.
}
\end{figure}

For the models with open boundary conditions,
we found the amplitudes to be on average 
higher than the vacuum case,
specially near the surface.
We have summarised in Fig.\ \ref{amplitudes.nfacb.MDI.sin4.ivac=9.clean.eps}
our calculations of 
the amplitudes of oscillations for the 
models with open boundary conditions given by
Eq.~(\ref{general}) and $n_2=0$, for a range of $n_1$ given by
$ 1 <n_1 < 150$. Here the dynamo parameters were 
$R_\alpha = -5.0, P_r = 0.7$, $R_\omega =50000$ and 
$f(\theta)= \sin^4 \theta\cos\theta$,
with $\alpha_r=1$ for $0.7 \leq r \leq 0.8$,
with cubic interpolation to zero at $r=r_0$ and $r=1$.
As can be seen, the amplitudes grow at all depths
with increasing $n_1$ and saturate around $n_1 \sim 50$.
As an example, the models with $n_1$ values around  $n_1=25$,
(corresponding to 
Fig.~\ \ref{ivac=7.velocity_radial_latitude=25.eps}, with STF), have 
amplitudes that are
more than double those found above
with vacuum boundary conditions,
down to the level $r_0= 0.88$.

\section{Discussion}
We have made a detailed study of the effects of 
the boundary conditions on the dynamics in
the solar convection zone, by employing 
various forms of outer boundary conditions.

In all the models considered here
(as well as other results not reported), we find that in 
near-critical and moderately supercritical
dynamo regimes the torsional oscillations extend all
the way down to the bottom of the
CZ. In this way our results, taken altogether, demonstrate
that such penetration is extremely robust
with respect to
all the changes we have considered both to the boundary conditions,
 and the  dynamo parameters such as
the dynamo and Prantdl numbers, in addition to variations in
the model ingredients
such as the $\alpha$ and $\eta$ profiles and
the rotation inversion.

We deduce, that if our dynamo model (which is basically
a standard mean field dynamo) has any validity, then
observers should expect to find that the solar torsional oscillations
penetrate to the tachocline.
However, given the significant uncertainties that
still remain in helioseismic measurements, especially
the limited temporal extent of the data available,
this issue may not be resolvable at present
(see e.g. Vorontsov et al. 2002).

In all cases we have found 
supercritical dynamo regimes 
with spatiotemporal fragmentation for
a range of, but not all, dynamo parameters. 
This, together with our previous work, shows that fragmentation occurs
with a variety of forms of $\alpha$ (and also that it is not confined
to a particular inversion for the solar angular velocity).
For still more supercritical dynamo regimes we find
a series of spatiotemporal fragmentations, leading eventually
to spatiotemporal chaos, i.e. 
disappearance of coherence in the dynamo regime.

These results are of potential importance in interpreting
the current observations, especially given their difficulty in
resolving the dynamical regimes near the bottom of the convection zone.
However given the variety of dynamical behaviour possible theoretically
near the bottom of the CZ, we cannot comment definitively on the
reported $1-3$ yr oscillations.

Finally, given the observational importance of the
amplitudes of the torsional oscillations, 
we have made a comparative study of their magnitudes
as a function of the boundary conditions.
We found that on average the amplitudes are smaller for 
the models with vacuum boundary conditions than for those with open boundary conditions. 
An important ingredient that our model omits
and which seems bound to have an effect on
the amplitudes of torsional oscillations
throughout the CZ, is that of density stratification.
We intend to return to this issue in a future publication.
 
\begin{acknowledgements}
RT benefited from UK Particle Physics and Astronomy Research Council
Grant PPA/G/S/1998/00576. EC and DM acknowledges the hospitality of the 
Astronomy Unit, QM.
\end{acknowledgements}


\begin{thebibliography}{}
\bibitem{antiaetal2000}

Antia H.M.,  Basu S., 2000, ApJ, 541, 442

\bibitem{covasetal2000a}
Covas E., Tavakol R., Moss D. \& Tworkowski A., 2000a, A\&A, 360, L21

\bibitem{covasetal2000b}
Covas E., Tavakol R. \& Moss D., 2000b, A\&A, 363, L13

\bibitem{covasetal2001c}
Covas E., Tavakol R. \& Moss D., 2001a,
A\&A, 371, 718

\bibitem{covasetal2001d}
Covas E., Tavakol R., Vorontsov, S. \& Moss, D., 2001b,
A\&A, 375, 260

\bibitem{covasetal2001e}
Covas E., Moss, D., \& Tavakol R., 2002,
Erratum, A\&A, to appear


\bibitem{howardetal1980}
Howard R. \& LaBonte B. J., 1980, ApJ Lett. 239, 33

\bibitem{howeetal2000a}
Howe R., et al., 2000a, ApJ Lett., 533, 163

\bibitem{howeetal2000b}
Howe R., et al., 2000b, Science, 287, 2456

\bibitem{kitchatinovetal2000}
Kitchatinov L.L., Mazur M. V. \& Jardine M., 2000, A\&A, 359, 531

\bibitem{kosovichevetal1997}
Kosovichev A. G. \&  Schou J., 1997, ApJ, 482, 207

\bibitem{}
Moss D., Mestel L. \& Tayler R.J., 1990, MNRAS, 245, 550

\bibitem{mossetal2000}
Moss D. \&  Brooke J., 2000, MNRAS, 315, 521

\bibitem{rudigeretal1995}
R\"udiger R. \& Brandenburg A., 1995, A\&A, 296, 557

\bibitem{schouetal1998}
Schou J., Antia H. M., Basu S. et al., 1998, ApJ, 505, 390

\bibitem{snodgrass1985}
Snodgrass H. B., Howard R. F. \& Webster L. 1985, Sol. Phys., 95,
221

\bibitem{mm}
Tworkowski A., Tavakol R., Brooke J.M., Brandenburg A., Moss D. \& Tuominen, I.,
1998, MNRAS, 296, 287
\bibitem{}
Vorontsov, S., Tavakol, R., Covas, E. \& Moss, D., (2002)
`Solar cycle variation of the solar internal rotation:
heleioseismic inversion and dynamo modelling',
To appear in `Proceedings of Granada Workshop
on `The evolving Sun and its influence on planetary
environments', ASP (Astronomical Society of the Pacific) Conference Series,
A. Gimenez, E. Guinan and B. Montesinos (eds), astro-ph/0201422, also
available at http://www.eurico.web.com
\end{thebibliography}
\end{document}